\documentclass[pdfusetitle,aps,reprint,prd,twocolumn,superscriptaddress,nofootinbib]{revtex4-1}

\usepackage{amssymb}
\usepackage{amsmath}
\usepackage{physics}
\usepackage{epsfig}
\usepackage{breakurl}
\usepackage{float}
\usepackage{multirow}
\usepackage{array}
\usepackage{bm}
\usepackage{color}
\usepackage{tikz}
\usepackage[utf8]{inputenc}
\usepackage{braket}
\usepackage{graphicx}
\usepackage{footmisc}
\usepackage{tensor}
\usepackage[normalem]{ulem}
\usepackage{dblfloatfix}
\usepackage[T1]{fontenc}
\usepackage{array}
\usepackage{booktabs}
\usepackage{multirow}
\usepackage{amstext}
\usepackage{multirow}

\newcommand{\M}{{\cal M}}
\usetikzlibrary{decorations.pathreplacing}



\usepackage[force]{feynmp-auto}

\xdefinecolor{mylinkcolor}{rgb}{0,0,0.7}
\usepackage[
	bookmarksnumbered, bookmarksopen, bookmarksopenlevel=2,
	breaklinks=true,
	colorlinks=true, filecolor=mylinkcolor, citecolor=mylinkcolor,
	linkcolor=mylinkcolor, urlcolor=mylinkcolor, menucolor=mylinkcolor,
]{hyperref}
\usepackage{bookmark}

\usepackage{cleveref}
\crefformat{section}{\S#2#1#3} 
\crefformat{subsection}{\S#2#1#3}
\crefformat{subsubsection}{\S#2#1#3}

\usepackage{colortbl}
\definecolor{blue2}{cmyk}{1, 0.1, 0.1, 0}

\definecolor{pyBlue}{RGB}{31, 119, 180}
\definecolor{pyRed}{RGB}{214, 39, 40}
\definecolor{pyGreen}{RGB}{44, 160, 44}
\definecolor{pyBlue2}{RGB}{0, 111, 237}
\definecolor{pyRed2}{RGB}{224, 52, 36}

\definecolor{summersky}{cmyk}{0.71,0.33,0,0.5}
\definecolor{flamingo}{cmyk}{0,0.51,0.71,0.5}
\definecolor{rp}{cmyk}{0.2, 1, 0.6, 0}
\definecolor{pacificblue}{cmyk}{0.95,0.3,0, 0.5}
\definecolor{gray60}{cmyk}{0.4,0.4,0,0.8}

\renewcommand{\]}{\right]}

\renewcommand{\vec}[1]{\mathbf{#1}}

\usetikzlibrary{decorations.markings}
\usepackage{cleveref}
\crefformat{section}{\S#2#1#3}
\crefformat{subsection}{\S#2#1#3}
\crefformat{subsubsection}{\S#2#1#3}

\makeatletter
\def\simgt{\mathrel{\lower$\frac{5}{2}$pt\vbox{\lineskip=0pt\baselineskip=0pt
           \hbox{$>$}\hbox{$\sim$}}}}
\def\simlt{\mathrel{\lower$\frac{5}{2}$pt\vbox{\lineskip=0pt\baselineskip=0pt
           \hbox{$<$}\hbox{$\sim$}}}}

\makeatother

\def\spa#1.#2{\left\langle#1\,#2\right\rangle}
\def\spb#1.#2{\left[#1\,#2\right]}
\def\sand#1.#2.#3{%
\left\langle#1{\vphantom1}\right|{#2}\left|#3\right]}%
\def\sandmp#1.#2.#3{%
\left\langle#1{\vphantom1}\right|{#2}\left|#3\right]}%
\def\sandpm#1.#2.#3{%
\left[#1{\vphantom1}\right|{#2}\left|#3\right\rangle}%
\def\sandmm#1.#2.#3{%
\left\langle#1{\vphantom1}\right|{#2}\left|#3\right\rangle}%
\def\sandpp#1.#2.#3{%
\left[#1{\vphantom1}\right|{#2}\left|#3\right]}%

\renewcommand{\imath}{\mathrm{i}}

\newcommand{\be}{\begin{equation}}
\newcommand{\ee}{\end{equation}}

\newcommand{\eqq}[1]{Eq.~(\ref{#1})}

\def\S{{\mathbb S}}

\allowdisplaybreaks

\begin{document}

\title{The Rise of Linear Trajectories}
\author{Yu-tin Huang}
\affiliation{Department of Physics and Center for Theoretical Physics, National Taiwan University, Taipei 10617, Taiwan}
\affiliation{Physics Division, National Center for Theoretical Sciences, Taipei 10617, Taiwan}
\affiliation{Max Planck{-}IAS{-}NTU Center for Particle Physics, Cosmology and Geometry, Taipei 10617, Taiwan}
\author{Sara Ricossa}

\author{Francesco Riva}

\affiliation{Départment de Physique Théorique, Université de Genève,
24 quai Ernest-Ansermet, 1211 Genève 4, Switzerland}
\author{Jie-Da Tsai }
\affiliation{Department of Physics and Center for Theoretical Physics, National Taiwan University, Taipei 10617, Taiwan}
\email{\;}

\email{}

\begin{abstract}
In this letter, we consider constraints on the low-energy spectrum of amplitudes with higher-spin exchange. Assuming unitarity, crossing symmetry, and super-convergent high energy behavior, reminiscent of the scattering of spin-1 and spin-2 massless helicity states, 
we demonstrate that the spectrum that maximizes the leading higher spin couplings of the second and third resonances is consistently given by a linear trajectory. Furthermore, for gravitational theories, the optimal spectrum is the linear trajectory defined by the mass and spin of the graviton and the lightest spin-4 resonance.
\end{abstract}

\maketitle

\section{Introduction}
Motivated by the success of the numerical conformal bootstrap, where unitarity and crossing symmetry are recast as a convex optimization problem~\cite{Rattazzi:2008pe}, substantial progress has been made in adapting semi-definite programming techniques to the S-matrix bootstrap~\cite{Paulos:2017fhb, Hebbar:2020ukp, Caron-Huot:2020cmc, Caron-Huot:2021rmr, Albert:2022oes}.
Within this framework, unitarity and analyticity constraints translate into sharp bounds on Wilson coefficients of effective field theories (EFTs), and on the couplings of low-lying resonances~\cite{Adams:2006sv,Arkani-Hamed:2020blm,Bellazzini:2020cot,Tolley:2020gtv,Cheung:2016yqr,deRham:2018qqo,Zhang:2018shp,Englert:2019zmt,Bellazzini:2019bzh,Remmen:2019cyz,Huang:2020nqy,Sinha:2020win,Caron-Huot:2021rmr,Bern:2021ppb,Chiang:2021ziz,Henriksson:2021ymi,Caron-Huot:2022ugt,Albert:2022oes,Ma:2023vgc,Li:2023qzs,Albert:2023seb,Albert:2024yap,Fernandez:2022kzi,Bellazzini:2023nqj,Bertucci:2024qzt,Berman:2024kdh,Dong:2024omo,Berman:2024wyt,Berman:2024eid,Pasiecznik:2025eqc,Bellazzini:2025shd}. 
These developments open the tantalizing prospect of isolating the minimal set of physical principles under which the string-theory S-matrix emerges as the unique, self-consistent solution.


In the early  S-matrix program, the  zero-width approximation and channel duality played a crucial  role in leading to the  Veneziano amplitude~\cite{Veneziano:1968yb} and reproduce Regge asymptotics~\cite{Dolen:1967jr}. 
In this approximation, the S-matrix is ``tree-level'' and the exchanged states are  stable particles of different spin and mass. A spectrum involving higher-spin states is infinite and highly constrained.  For example, the gap above the massive spin-2~\cite{Bellazzini:2023nqj}, as well as  higher spin states~\cite{Albert:2024yap,Bellazzini:2025shd}, must be finite. Furthermore,  the mass of the lightest spin-$\ell$ state must be smaller than that of the lightest spin-$(\ell{+}2)$ state, and satisfy   convexity properties~\cite{Berman:2024eid, Berman:2024kdh, Berman:2024owc}.

In this work,  we study the low lying  spectrum that can be exchanged at tree-level in scalar amplitudes, consistent with unitarity, crossing symmetry and twice-subtracted dispersion relations at fixed momentum transfer, which apply to gapped, as well as gravitational theories~\cite{Haring:2022cyf}. In this context, it was recently shown that  the additional high-energy (UV) assumption of ultrasoftness   and patterns of zeros, uniquely fix the four-point amplitude  to that of Veneziano and Virasoro--Shapiro~\cite{Cheung:2025tbr}. Here, we instead focus on low-energy (IR) properties of the amplitude, and ask what further assumptions lead to string-like configurations.  

We  focus on the coupling of the lightest spin-$\ell$ state, and search for spectra that  maximize it.\footnote{This setup is immune to degeneracies associated with linear combinations of string theory amplitudes observed in  studies of extremal EFT coefficients~\cite{Haring:2023zwu,Gadde:2025fil}.} 
The couplings of spin-0 and spin-2 states cannot be bounded. This  can be traced to the existence of   amplitudes that satisfy bootstrap constraints but, in addition to  spin-0 or spin-2 states, have an unphysical accumulation point at $m_\infty$~\cite{Albert:2022oes},
\begin{equation}\label{scalarinftyamp}
    \M(s,t)= \frac{\lambda}{2}\frac{G_{\ell}^D\left(1{+}\frac{2t}{m^2}\right)}{m^2-s}\frac{m_{\infty}^2}{m_{\infty}^2-u}+{\rm perm.}\,,
\end{equation}
  with $\ell=0,2$ and $G^D_{\ell}$  the Gegenbauer polynomial in $D$-dimensions.
  
  Here we circumvent this by instead considering $-2$ and $0$ subtractions. Such a setup can be viewed as realizing maximal/half-maximal supersymmetry (with 32/16 supercharges), echoing  improved subtractions familiar for scattering of spin-2 and spin-1 helicity states in $D=4$~\cite{Caron-Huot:2022ugt,Henriksson:2021ymi}.\footnote{Bounds on interactions of the first massive state show features of both maximal SUSY~\cite{Albert:2024yap} and graviton amplitudes in $D$=4~\cite{Pasiecznik:2025eqc}. } In this work we focus on $D=10$, where type-IIA/B string theory operates, but we expect the result to be more general.

 Remarkably,  by studying the couplings of the lightest few states as a function of masses, we find that \emph{a linear spectrum consistently maximizes the couplings} and singles out a linear leading Regge trajectory. In the presence of gravity, the  maximal couplings correspond to a linear trajectory with zero intercept,  passing through the graviton pole, marking  a sharp distinction to non-gravitational theories.

These results are reminiscent of studies in large-$N$ QCD, where the functional that yields the maximal coupling of the  spin-2 resonance  in the presence of a lighter spin-1 state exhibits a single (almost) linear trajectory~\cite{Albert:2022oes, Albert:2023jtd,Albert:2023seb}. However, it is known that single trajectories cannot be self consistent~\cite{Eckner:2024pqt}, hence the  interpretation of such extremal spectra remains an open question. Here, by directly studying the coupling of a spin-$\ell$ resonance as a function of the leading spectrum, we are able to put  the emergence of linear trajectories on a stronger footing, independently of the existence of sub-leading trajectories.

\section{Setup}
We consider the 2$\rightarrow$2 scattering amplitude of massless scalars $\M(s,t)$, where $s=(p_1{+}p_2)^2$ and $t=(p_1{-}p_4)^2$. Assuming that the amplitude involves only tree-level exchanges, $\M(s,t)$ is a meromorphic function with simple poles reflecting $s,t$ and $u={-}s{-}t$ channel exchanges. Near these poles, the amplitude behaves as,
\begin{equation}\label{eq:resonances}
\M(s,t)\bigg|_{s\rightarrow m_i^2}=-\frac{ \sum_{\ell} \lambda_{i,\ell} G^D_{\ell}(1{+}\frac{2t}{m_i^2})}{s-m_i^2}\,,
\end{equation}
 with $\lambda_{i,\ell}\geq0$, by unitarity. The amplitude is bounded by $s^2$ in the Regge limit, i.e.\
\begin{equation}
\lim_{s\rightarrow \infty} \frac{\M(s,t)}{s^2}=0,\quad \text{fixed}\;t\,,
\end{equation}
which is guaranteed 
in gapped theories, 
and follows from the emergence of classical GR at large impact parameter in gravitational theories~\cite{Haring:2022cyf}. 

For 10-dimensional  theories with $\mathcal{N}=2$ maximal (32 super-charges) and $\mathcal{N}=1$ half-maximal (16 super-charges) supersymmetry (SUSY), the amplitude takes the form 
\begin{equation}\label{eq: SUSYAmp}
\M(s,t)=\delta^{8\mathcal{N}}(Q)f_{\mathcal{N}}(s,t)\,
\end{equation}
with $\delta^{8\mathcal{N}}(Q)$ the fermionic delta function that solves the SUSY Ward identities at four-point (see~\cite{Albert:2024yap} for a review).
Since $\delta^{8\mathcal{N}}(Q)\sim s^{2\mathcal{N}}$, this implies an improved UV behavior for $f_{\mathcal{N}}$, 
\begin{eqnarray}
\lim_{s\rightarrow \infty} f_{\mathcal{N}}(s,t)s^{2(\mathcal{N}{-}1)}=0,\;\text{fixed}\;t\,.
\end{eqnarray}
This is reminiscent of graviton ($\ell$=2) and photon ($\ell$=1) scattering in four dimensions, where, for specific helicity configurations, the amplitude is proportional to $s^{2\ell}$ and leads to an identical improved UV behavior~\cite{Caron-Huot:2022ugt}. Indeed, for $\mathcal{N}=2$, the massless multiplet contains the graviton in addition to the scalar, while for $\mathcal{N}=1$ it contains the photon.

\vspace{5mm}
In this context, we study fixed $t=-p^2$ dispersion  relations. These relate integrals, 
\begin{equation}\label{eq:archdeft} 
    a^\mathcal{N}_n(p^2) \equiv
    \oint\frac{{\rm d} s}{2\pi i \, s}\frac{f_\mathcal{N}(s,-p^2)}{[s \, (s-p^2)]^{n/2}} \,,\quad
    n\geq0\,,
\end{equation}
along a circle of radius $M^2+t/2$ around $-t/2$, and integrals along the real $s$ axis,
\begin{align}\label{eq:fixedt}
    a^\mathcal{N}_n\left(p^2\right) &= \sum_{\ell\text{ even}}n_\ell^D\int_{M^2}^\infty\frac{{\rm d} s}{2\pi}\frac{(2s{-}p^2)\rho_\ell(s)s^{\frac{4{-}D}{2}}}{\[s(s{-}p^2)\]^{n/2{+}1}}
        G^D_{\ell}(s,{-}p^2)\\\nonumber
        &\equiv \left\langle\frac{2s{-}p^2}{\[s(s{-}p^2)\]^{n/2{+}1}}
        G^D_{\ell}(s,{-}p^2)\right\rangle,
\end{align}
for even $n\,{\geq}\,{-}2$ or $n\,{\geq}\,0$ for $\mathcal{N}\,{=}\,2,1$, respectively. 
Here, $\rho_\ell(s)$ is the spectral density, positive by unitarity. For narrow resonances at mass $m_i$, $n_\ell^D\rho_\ell(s)=2\pi\lambda_{i,\ell}s^{\frac{D{-}4}{2}}\delta\left(s-m_i^2\right)$, see \eqq{eq:resonances}.

Beautifully, the spectral density must further enable a description of the amplitude as a sum of dual $t$-resonances, implying a series of null constraints on the spectral density itself~\cite{Caron-Huot:2020cmc,Tolley:2020gtv}. These can be constructed in the forward limit via a double contour integral~\cite{Albert:2023jtd}:
\begin{eqnarray}
\chi_{n,k} &=& \text{Res}_{t=0} \Bigg[ 
\left(\frac{1}{s^{n{-}k-1} t^{k+1}} {-} \frac{1}{t^{n{-}k-1} s^{k{+}1}} \right)\nonumber \\
&& \quad {-} \big( s \to {-}s{-}t \big)
\Bigg]\mathcal{P}_\ell \!\left( 1 {+} \frac{2t}{s} \right)\,,\label{eq:NCS}
\end{eqnarray}
where $\langle \chi_{n,k}\rangle = 0$ for $n = 3, 4, 5, \ldots$, and, $k = 0, 1, \ldots, (n-2)/2$ in the case without gravity -- with gravity, $n=4,5,6,\ldots$ instead.\footnote{
In the presence of loop effects from massless particles, only some of these null constraints survive; see~\cite{Bellazzini:2021oaj,Beadle:2024hqg,Beadle:2025cdx,Chang:2025cxc}.}

\vspace{5mm}

At  energies lower than the mass of the exchanged massive states, the amplitude is well described by an EFT with, 
\begin{eqnarray}\label{eq: LEDef}
f^{\rm IR}_{1}(s,t)&=&-8\pi G_N\left(\frac{1}{s}+\frac{1}{t}+\frac{1}{u}\right)+g_0{+}g_2\sigma_{2}{+}\cdots,\nonumber\\ f^{\rm IR}_{2}(s,t)&=&\frac{8\pi G_N}{stu}{+}g_0{+}g_2\sigma_{2}{+}\cdots
\end{eqnarray}
where $\sigma_{2}=s^2{+}t^2{+}u^2$ and $\cdots$ representing higher derivative terms. For $\mathcal{N}=1$,  $g_0$ and $g_2$ can be identified with  the Wilson coefficients of the operators $F^4$ and $D^4F^4$ in the photon sector, while for $\mathcal{N}=2$, it is  maximal supergravity.
Substituting \eqq{eq: LEDef} into \eqq{eq:archdeft}, we find for example,
\begin{equation}
a^2_{-2}=\frac{8\pi G_N}{p^2},\quad a^1_{0}=g_0{+}2g_2p^4{+}\cdots. 
\end{equation}

In the rest of this article we shall be interested in how the null constraints \eqq{eq:NCS} shape spectra with narrow resonances. The EFT coefficients in \eqq{eq: LEDef}---which have been the focus of most of the EFT bootstrap literature---will only serve us as normalization factors.

\section{Zero-Subtractions}

We begin  the 0-subtracted scenario, where the amplitude  \emph{must} include the exchange of infinitely many resonances.
An example of a UV complete amplitude satisfying these properties 
is the open superstring completion of the DBI action~\cite{Metsaev:1987qp}, where the four-point amplitude is given by,
\begin{equation}\label{eqDBI}
f_1(s,t)={-}\frac{\Gamma\left({-}t\right)\Gamma\left({-}u\right)}{\Gamma\left(1{+}s\right)}
{-}\frac{\Gamma\left({-}s\right)\Gamma\left({-}u\right)}{\Gamma\left(1{+}t\right)}
{-}\frac{\Gamma\left({-}s\right)\Gamma\left({-}t\right)}{\Gamma\left(1{+}u\right)}\,,
\end{equation}
and where we normalize the string scale $\alpha'=1$. This function has $s$-channel poles at $s=1,3,5,\cdots,{2k}{+}1$, where at $s=2k{+}1$ lies the highest-spin  $\ell=2k$ state (the factor of $2$ comes from the $\delta^{8}(Q)$ in Eq.~(\ref{eq: SUSYAmp})).

 
We assume the existence of a finite number of  isolated narrow resonances at scales below $M$.
 Above $M$, on the other hand, we remain agnostic about the spectrum and refer to this region as the UV.\@ We aim to bound the couplings of the resonances normalized by the low-energy data, such as the EFT Wilson coefficients or $G_N$. Note, however, that since the DBI amplitude is already self-consistent without a graviton pole, there can not be any bound when normalized against $G_N$. Thus, we normalize the couplings with respect to $g_0$ and set $G_N=0$ for simplicity.   

\subsection{Two-states system}
We begin with an ansatz for the spectrum which involves two sets of isolated states with masses $0<m_1^2<m_2^2\leq M^2$ and spins $\ell_1=0$, $\ell_2\leq 2$, in addition to the massless scattered particles. 
For this  setup, the precise ansatz that enters \eqq{eq:fixedt} is, 
\begin{eqnarray}
\tilde{\rho}^{2{\rm state}}_\ell(s)&=&\delta_{\ell,0}\lambda_{1,0}\delta(s{-}m^2_1){+}\sum_{\ell'=0}^2\delta_{\ell,\ell'}\lambda_{2,\ell'}\delta(s{-}m^2_2)\nonumber\\
&{+}& \tilde{\rho}_\ell(s)\theta(s-M^2)\,,\label{eq: RhoDef2}
\end{eqnarray}
where $\tilde{\rho}_\ell(s)\,{\equiv}\,n^D_\ell\,s^{\frac{4{-}D}{2}}\rho_\ell(s)/2\pi$, and we use $\lambda_{i,\ell}$ to represent the coupling of the $i$-th state with spin $\ell$.

\begin{figure}[h]
\begin{center}
\includegraphics[width=\linewidth]{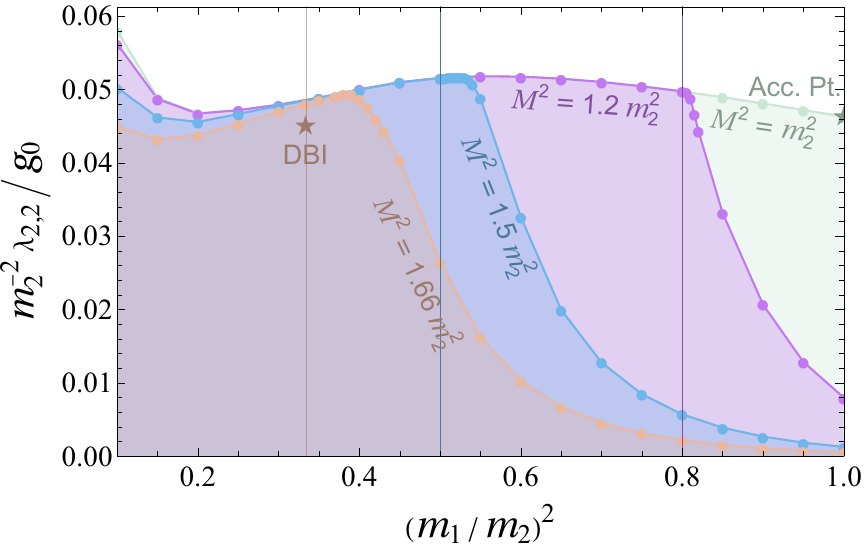}
\end{center}
\caption{\footnotesize \emph{Maximization of $\lambda_{2,2}/g_0m_2^2$ in a system with two resonances below the threshold $M$, for different values of $m_1$ and $M$. Vertical lines at critical values $m_1=m_{1,c}$ saturating \eqq{lineq}. All our results are in $D=10$ dimensions. See  App.~\ref{app:sdpb1} for details on numerics (the small rise at $m_1\to 0$ is a numerics artifact). Orange star for DBI.\@ Green star for accumulation-point amplitude:
$\left[(s{-}M^2)(t{-}M^2)(u{-}M^2)\right]^{{-}1}$.}}\label{fig:2state}
\end{figure}

Fixing $m_2$ and $M^2$, we then maximize the coupling $\lambda_{2,2}$---more precisely, the dimensionless ratio $\lambda_{2,2}/g_0 m_2^2$---as a function of $m_1$. 
We repeat this  for different values of $M$, and
show the results in Fig.~\ref{fig:2state}. We observe that as we increase $m_1$, the maximal value tracks an approximate plateau until an $M$-dependent critical value, after which it sharply drops to zero.

\begin{figure}[h]
\begin{center}
\includegraphics[width=\linewidth]{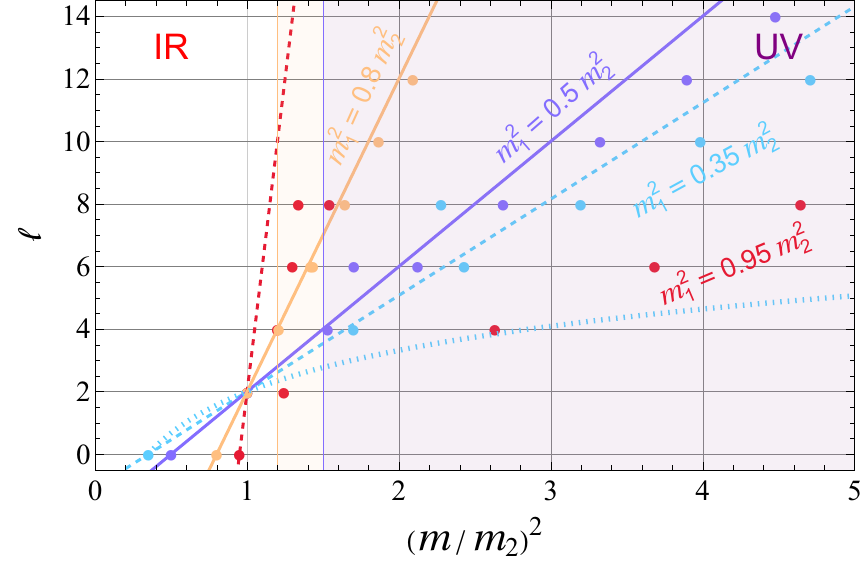}
\end{center}
     \caption{\footnotesize \emph{
     Spectra that maximize the coupling of Fig.~\ref{fig:2state} (shown only states with $\lambda_{i,\ell}\geq 10^{-5}g_0m_1^2$). The yellow shaded area represents the UV, for $M^2=1.2m_2^2$. For this value of $M$, the solid yellow line is extremal  
       $m_1^2= m_{1,c}^2=0.8m_2^2$, and has
     linear trajectories that emerge and persist in the UV,
     just like the sub-extremal light and dark blue lines at $m_1^2<m_{1,c}^2$. Lines are plotted with slopes fixed by the two lighter states. In red,  $m_1>m_{1,c}$ for which the linear trajectory is not  allowed by the large $M$, and does not appear in the UV.\@ With higher $M^2=1.5m_2^2$  the blue solid line becomes extremal $m_{1,c}^2=0.5 m_2^2$ (UV region in shaded blue). The light blue spectrum is unaffected by these choices of $M$. Light blue dotted: the exponential spectrum saturating the analytic bounds~\cite{Berman:2024kdh,Berman:2024owc}.
}}\label{fig:2statem2}
\end{figure}

The sharp departure from the plateau can be consistently understood as due to coupling maximization by a linear spectrum in the spin/mass-squared plane. Since the spin-4 state can only occur at $M^2$ or higher, this implies that in order for there to be a linear spectrum, built from $(m_1^2,\ell_1{=}0)$, $(m_2^2,\ell_2{=}2)$ and $(m_3^2,\ell_3{=}4)$, where the putative $m_3^2\geq M^2$,  the following  inequality must be satisfied,
\begin{equation}\label{lineq}
m_1^2\leq 2m_2^2{-}M^2\equiv m^2_{1,c}\,.
\end{equation}
We call $m^2_{1,c}$ the critical value that saturates this inequality -- see vertical lines in  Fig.~\ref{fig:2state}, as well as  solid lines in Fig.~\ref{fig:2statem2}.
The sharp drops in Fig.~\ref{fig:2state} are in the proximity of $m^2_{1,c}$.

This argument is further strengthened by the observation that on the plateau, the maximal coupling and the corresponding spectrum are identical across different values of $M$: once a linear spectrum is possible, it is selected by the maximization.

In Fig.~\ref{fig:2statem2} we show the optimal spectra on the plateau: different lines correspond to different linear trajectories defined by different values of $m_1$. The light/dark blue lines allow  linear trajectories for cutoffs at both $M^2=1.2(1.5) \,  m_2^2$, and indeed their spectrum is unaffected by these choices of $M$.  
 Here, besides showing states lighter than $M$, we also extract the UV  spectra
 by studying the optimal functional in the semi-definite optimization problem. We only show states with couplings $\lambda_{i,\ell}\geq 10^{-5}g_0m_1^2$.  The approximate  linear trajectories persist in the UV.\@

Note that Refs.~\cite{Berman:2024kdh,Berman:2024owc} found analytic constraints on meromorphic spectra from studying dispersion relations with $n{\to}\infty$
subtractions. In our setup, these would read $(M/m_1)^2\leq(m_3/m_1)^2\leq (m_2/m_1)^4$, or concretely $m_1^2\leq m_{1,a}^2\equiv  (5/6,2/3,0.6)\,  m_2^2$ for $M^2=(1.2,1.5,5/3)  \,m_2^2$. One might wonder if the  sharp fall-off is really a reflection of these analytic bounds, rather than our hypothesis. Comparing the critical mass for the analytic bounds, $m_{1,a}^2$, with the linear trajectory values,  $ m_{1,c}^2 = (0.8,0.5,1/3)\,m_2^2$, we see that they are distinct, and the fall-off starts closer to $m_{1,c}^2$ than $m_{1,a}^2$. Furthermore, the extremal spectrum near the fall-off is in fact drastically different from the spectrum that saturates the analytic bounds, which is given by  $m_{1+\ell/2}^2\,{=}\,m_1^2(m_2/m_1)^\ell$. We illustrate this in Fig.~\ref{fig:2statem2} by comparing the  dashed and dotted blue curves. The analytic bounds are therefore not optimal yet, but it would be nice to see if they can be further improved.

On the other hand, the analytic results imply sharp bounds on the mass-spectrum, a feature which is not directly visible in our finite-$n$ results. We provide a discussion of 
$n$-convergence in  Appendix~\ref{app:conv}.


\subsection{Three-state system}
To further establish the prominence of linear trajectories, we  bring $m^2_3$ into the IR spectrum where we now have $(m_1^2,\ell_1{=}0)$, $(m_2^2,\ell_2{\leq}2)$ and $(m_3^2,\ell_3{\leq}4)$, with $m_1^2\leq m_2^2 \leq m_3^2 \leq M^2$. The ansatz for the spectral density now takes the form,
\begin{align}\label{eq: RhoDef3}
\tilde{\rho}^{3{\rm state}}_\ell(s)=&\delta_{\ell,0}\lambda_{1,0}\delta(s{-}m^2_1){+}\sum_{\ell'=0}^2\delta_{\ell,\ell'}\lambda_{2,\ell'}\delta(s{-}m^2_2)\nonumber\\
&{+}\sum_{\ell''=0}^4\delta_{\ell,\ell''}\lambda_{3,\ell''}\delta(s{-}m^2_3){+} \tilde{\rho}_\ell(s)\theta(s-M^2)\,.
\end{align}

\begin{figure}[ht]
\begin{center}
\includegraphics[width=\linewidth]{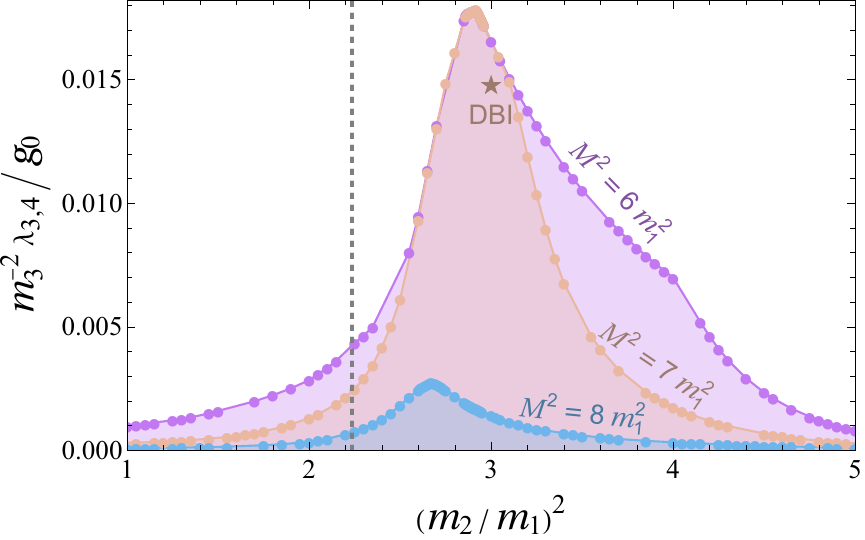}
\end{center}
\caption{\footnotesize\emph{Maximal coupling of the $\ell=4$ state, as a function of $m_2$ for $m_3^2=5m_1^2$, and for different values of $M^2/m_1^2=6,7,8$, with $7$ the reference DBI value, and the DBI result denoted by a star. Analytic lower bound shown by the  dashed line~\cite{Berman:2024kdh,Berman:2024owc}.}}\label{fig:3state_muChange}
\end{figure}

In this setup, we search for maxima of the coupling of the spin-4 state $\lambda_{3,4}/g_0 m_3^2$, as a function of $m_2$.\footnote{In the  maximization  of the lighter state coupling $\lambda_{2,2}$, the states at $m_3$ 
can be effectively removed by the optimization procedure, thus reducing to the 2-state ansatz. 
 By instead maximizing the coupling to the last state before the cutoff, the lighter states are required not to vanish~\cite{Berman:2024kdh,Berman:2024owc}.} The results are shown in Fig.~\ref{fig:3state_muChange}, for $m_1^2=1$, $m_3^2=5$ and for different values of the cutoff $M$. The linear trajectory hypothesis is made manifest by the sharp peak located close to $m_2^2=(m^2_1{+}m^2_3)/2$. Furthermore, for values of $M$ that are compatible with higher states lying on the linear trajectory, i.e.\ an $\ell=6$ state at $m^2_4=(3m^2_3-m_1^2)/2\geq M^2$, this sharp peak is almost unchanged, as illustrated by the purple and orange curves in the figure. Instead for larger values of $M$ that do not allow for a linear trajectory, the coupling is suppressed -- see blue curve in Fig.~\ref{fig:3state_muChange}.

Away from the peak, the optimal spectrum is given by a linear trajectory with $(m_2^2,\ell_2=2)$, $(m_3^2,\ell_3=4)$ and putative $(m_4^2>M^2,\ell_4=6)$, at the expense of the states at $m_1$, which are substituted by sub-leading states at $m_2$. This is made manifest by the sharp kink on the purple curve at $m_2^2=4m_1^2$, which is the uppermost value for such a trajectory without $m_1$. This is  verified by direct inspection of the optimal spectra -- see Appendix~\ref{app:3st}.

Finally, in this case without gravity, our results depend only very weakly on the line's steepness and intercept, as long as a linear trajectory forms; see discussion in Appendix~\ref{app:3st}.

Our results have deeper implications than the exact shape of the maximal curve. The sharp drops to zero that we observe away from the linear trajectories, imply that 
linear trajectories provide the bulk of the contribution to the low-energy couplings, not just the maximal one. In this sense, linear spectra do not merely emerge as the limiting solutions of the optimization problem, but are a defining feature of all non trivial theories.

\section{Gravity and $-2$ Subtractions}
We now consider the case with gravity and maximal supersymmetry,  captured by $-2$ subtracted dispersion relation. In this case, a relevant UV completion is~\cite{Arkani-Hamed:2020blm},
\begin{equation}\label{eq: DeformAmp}
f_2(s,t) =\frac{{-}\Gamma({-}s)\Gamma({-}t)\Gamma({-}u)}{\Gamma(1{+}s)\Gamma(1{+}t)\Gamma(1{+}u)}\left(1{-}\epsilon\frac{stu}{(1{+}t)(1{+}u)(1{+}s)}\right)\,.
\end{equation}
This amplitude is unitary for $0\leq\epsilon\leq1$,\footnote{This amplitude is an example where the conditions in~\cite{Cheung:2025tbr} are met, including ultra-soft high-energy behavior, except for the pattern of zeros. } and reduces to the Virasoro--Shapiro amplitude for $\epsilon=0$. In addition to the graviton pole, this function has $s$-channel poles at $s=1,2,\cdots$, where at $s=k$ lies the highest-spin  $\ell=2k{-}2$ state.
\subsection{Two-states system}

\begin{figure}[h]
\begin{center}
\begin{tikzpicture}[>=stealth]
\node[anchor=north west,inner sep=0] (img) at (-1,-2.5) {\includegraphics[width=\linewidth]{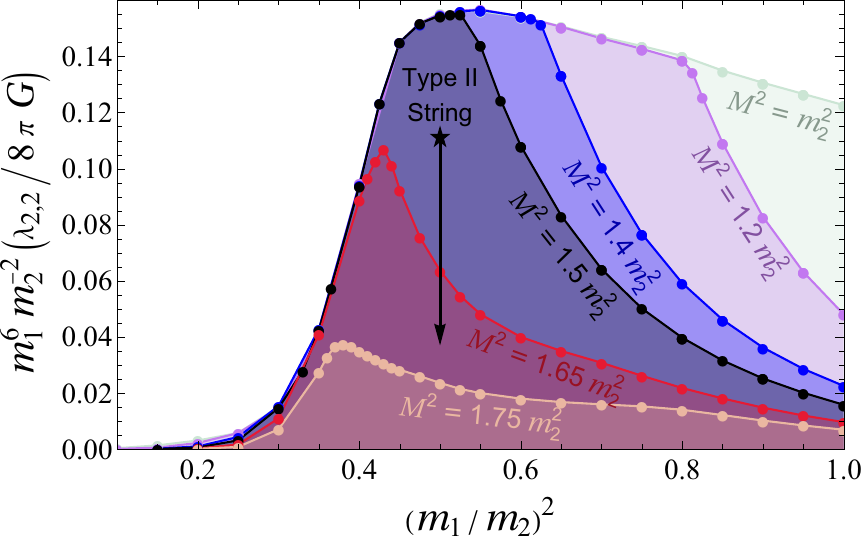}};
\label{2-state}
\begin{scope}[yscale=0.7]
\draw[dashed] (3.4,-3.58) -- (1.8,0);
\draw[dashed, blue] (4.22,-3.58) -- (2.16,0);
\draw[dashed, red] (2.2,-3.58) -- (1.26,0);
\draw[->,thick] (0,0) -- (7,0) node[right] {$(\frac{m}{m_2})^2$};
\draw[->,thick] (0,-3) -- (0,5) node[above] {$\ell$};

\foreach \y in {-2,2,4}
  \draw (-0.1,\y) -- (0.1,\y) node[left=4pt] {\y};

\draw[thick,red] (1.26,0) -- (5.94,4);
\draw[thick,blue] (2.16,0) -- (5.04,4);

\draw[black,thick] (0,-2) -- (5.4,4);

\fill[red!20,opacity=0.5] (5.94,0) -- (5.94,5) -- (7,5) -- (7,0) --cycle;
\fill[blue!20,opacity=0.5] (5.04,0) -- (5.04,5) -- (7,5) -- (7,0) --cycle;

\draw[red,dashed] (5.94,0) -- (5.94,5);
\draw[blue,dashed] (5.04,0) -- (5.04,5);
\draw[black,dashed] (5.4,-0.5) -- (5.4,5);
\draw[black,dashed] (0,4) -- (7,4);

\draw[dashed] (3.6,0) -- (3.6,2);
\coordinate (A) at (1.26, 0);
\coordinate (B) at (1.8, 0);
\coordinate (C) at (2.16, 0);
\coordinate (D) at (3.6, 2);
\end{scope}
\fill[red] (A) circle (2pt) ;
\fill[black] (B) circle (2pt);
\fill[blue] (C) circle (2pt) ;
\fill (D) circle (2pt) node[below,yshift=-1.45cm, xshift=0pt] {$1$};
\node[below,yshift=-0.05cm] at (5.04,0) {$1.4$};
\node[below,yshift=-0.4cm] at (5.4,0) {$1.5$};
\node[below,yshift=-0.05cm] at (5.94,0) {$1.65$};
\end{tikzpicture}
\end{center}
\caption{\footnotesize \emph{
TOP: A cartoon  of linear spectra,  for different choices of~$M^2$.  Blue/ black/red lines represent lower-than-critical/ critical/larger-than-critical slopes. BOTTOM: maximization of $\lambda_{2,2} m_1^6/8\pi G m_2^2$ in a system with two resonances below threshold, for different values of $m_1^2$ and $M^2$. The vertical black arrow correspond to the amplitude in Eq.~(\ref{eq: DeformAmp}), with the star indicating the $\epsilon=0$ case.
    }}\label{fig:2stateTheory}
\end{figure}
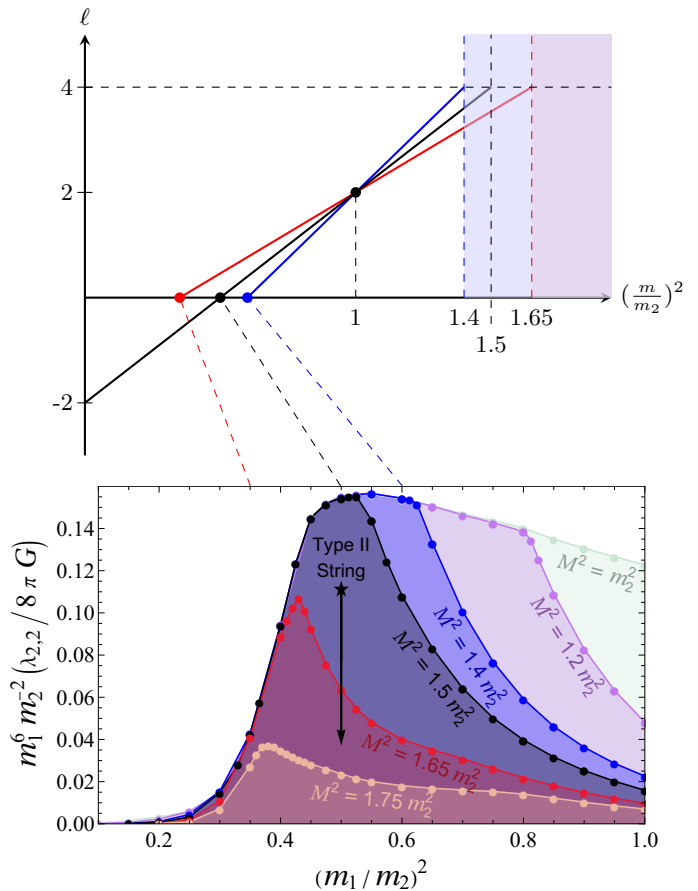

We start with two isolated states below the gap $M^2$ and maximize the spin-2 coupling $\lambda_{2,2}$ for different values of $m_1^2/m_2^2$. The result is shown in Fig.~\ref{fig:2stateTheory}. 

Compared to the 0-subtraction analysis in Fig.~\ref{fig:2state}, we see that here there is no plateau at $m_1^2 < m_2^2/2$, independently of $M^2$. Moreover, for $M^2 < 3 m_2^2/2$, 
the curves peak
at $m_1^2=m_2^2/2$
and are followed by
a slow-roll slope  at $ m_2^2/2 \leq m_1^2 \leq m^2_{1,c}$. Instead, for $M^2>3m_2^2/2$, the coupling is drastically suppressed. 
{We interpret this as follows:
\begin{itemize}
\item Similarly to the 0-subtracted case, coupling maximization occurs in association with a linear spectrum interpolating massive states at $(m^2_1,\ell=0)$, $(m^2_2,\ell=2)$ and $(m^2_3>M^2,\ell=4)$. This explains the  sharp drop-off at $m_1^2>m^2_{1,c}$.
\item In this case, the graviton trajectory, defined by the graviton pole and $(m^2_1,\ell=0)$, defines a critical slope, below which trajectories can no longer maximize the coupling -- we sketch this in the top part of Fig.~\ref{fig:2stateTheory}. Hence the drop-off for $m_1^2 < m_2^2/2$ and also for $M^2>3m_2^2/2$. Even though trajectories with slopes larger than the graviton trajectory are not suppressed as strongly, the maximum of $\lambda_{2,2}$ remains on the graviton trajectory.
\end{itemize}
}

Therefore, the graviton pole leaves a distinctive imprint, with  the graviton trajectory providing the optimal solution for maximizing the spin-2 coupling. The vertical line in Fig.~\ref{fig:2stateTheory} denotes the values spanned by Eq.~(\ref{eq: DeformAmp}). Maximization indeed selects the Virasoro--Shapiro amplitude. We  provide details on the numerical implementation and convergence  in Appendix~\ref{app:jedi_appendix}.

\subsection{Three-states system}
\begin{figure}[h]
\includegraphics[width=\linewidth]{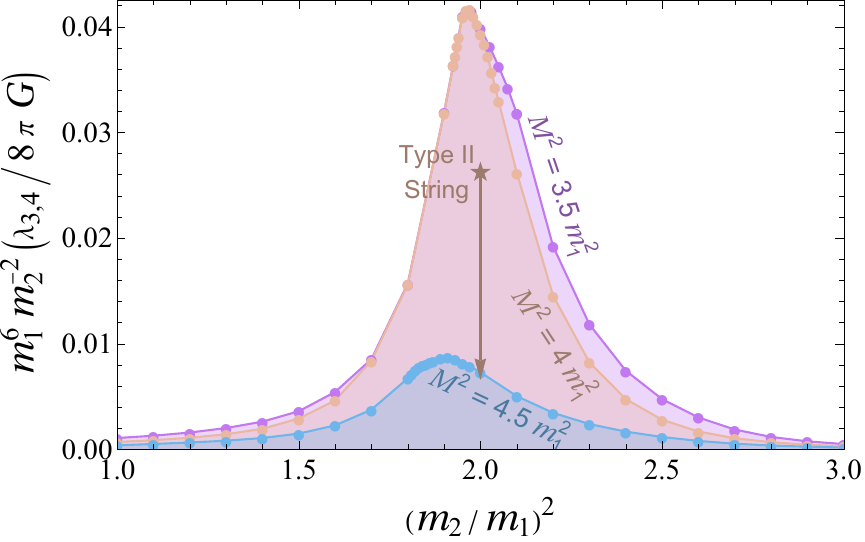}
 \caption{\footnotesize\emph{Maximum of $\lambda_{3,4} m_1^6/8\pi G m_3^2$, as a function of $(m_2/m_1)^2$, for $m_3^2=3 m_1^2$ and  different values of $M^2/m_1^2$.}
 }\label{grav_3state_m3}
\end{figure}

Further considering 3 resonances below threshold, we fix ($M^2, m_3^2, m_1^2$) and maximize $\lambda_{3,4}$ with respect to $m_2^2$. The result for $m_3^2=3m^2_1$ is shown in Fig.~\ref{grav_3state_m3}.  A narrow peak emerges at the critical value  $m_2^2 = (m_1^2 + M^2)/2$ which satisfies \eqq{lineq}, and also lies on the graviton trajectory. 
When $M^2$ falls below $4m^2_1$,  the graviton trajectory is no longer viable, and the peak drastically decreases. 

\newpage
\section{conclusion}
\vspace{-9pt}
In this letter, we conjecture that in the narrow-width approximation, maximizing the couplings of higher-spin states leads to linear trajectories. We test this conjecture by studying the leading spin-2 and spin-4 states in the scattering of half-maximal and maximal massless super-multiplets, which satisfy $0$ and $-2$ subtraction dispersion relations respectively. 

We find, moreover,  that theories with non-trivial couplings---and not just maximal ones---consistently occur when the low lying states can be aligned onto, or in the proximity of, a linear trajectory. When gravitational effects are turned on, the gravitational trajectory---where the leading higher spin-states are lined up with an intercept that matches the graviton pole---provides the optimal solution. 

In the traditional Regge theory program, linear Regge trajectories emerged as a global feature, arising from the requirement that an infinite tower of resonances collectively produce finite high-energy growth of scattering amplitudes, i.e.\ Reggeization.
By contrast, in our analysis, we do not assume the existence of an entire Regge tower apriori,  do we assume any extraordinary UV softness; yet, linear trajectories emerge.

We further believe that these results generalize qualitatively to twice subtracted dispersion relations without SUSY, where the role of the spin-0 resonance is taken by the first massive spin-4 state.

Finally, we have  focused on the leading trajectory, with the understanding that daughter trajectories must be present, given the inconsistency of a single linear trajectory. It will be interesting to understand which IR principles organize the daughter trajectories as well. Furthermore, the narrow-width approximation also implies that the massive higher-spin states are stable to leading order in perturbation theory, such that the amplitudes of their scattering are amenable to the standard bootstrap constraints. The study of such mixed systems can potentially sharpen the picture of which low energy spectra maximize higher-spin couplings.


\section{acknowledgments}
We are very grateful to Justin Berman, Li-Yuan Chiang, Lukas~Lindwasser, Francesco~Sciotti and He-Chen Weng for important discussions, and Brando Bellazzini, Justin~Berman and Leonardo Rastelli for enriching comments on the manuscript. 
S.R. and F.R. have been supported by the SNSF under grant no. 200021-205016. Y-t Huang and J-D Tsai is supported by Taiwan National Science and Technology Council grant 112-2628-M-002-003-MY3 and 114-2923-M-002-011-MY5.

\clearpage
\onecolumngrid{}
\appendix
\section{Half-Maximal SUSY Numerical Implementation}\label{app:sara_appendix}
\subsection{Semidefinite Optimisation}\label{app:sdpb1}
%

In Fig.~\ref{fig:3state_muChange}, we bound the ratio $m_3^{-2}\lambda_{3,4}/g_0$ as a function of
$m_2$.  
We use the dispersion relations~\eqref{eq:fixedt} at $p^2=0$ to derive the bound, starting from the following  equality for $g_0$,
\begin{equation}
    a g_0 = \sum_{i,\ell} \lambda_{i,\ell}\left(\frac{2a}{m_i^2} + \vb*{b}\cdot\vb*{\chi}\left(m_i^2,\ell\right)\right)
    + \ev{\frac{2a}{s} + \vb*{b}\cdot\vb*{\chi}(s,\ell)},
\end{equation}
where the sum runs over the states below threshold in  the ansatz \eqq{eq: RhoDef2}, 
 $\vb*{\chi}$ is a vector containing the null constraints, while $a$ and $\vb*{b}$ are real parameters of the optimization procedure.
To bound $\lambda_{3,4}$, we single it out and write the relation as,
\begin{align}
    a g_0 - \frac{\lambda_{3,4}}{m_3^2} = &\lambda_{3,4}\left(-\frac{1}{m_3^2} + \frac{2a}{m_3^2} + \vb*{b}\cdot\vb*{\chi}\left(m_3^2,4\right)\right)
    +\sum_{\left(i,\ell\right)\neq(3,4)} \lambda_{i,\ell}\left(\frac{2a}{m_i^2} + \vb*{b}\cdot\vb*{\chi}\left(m_i^2,\ell\right)\right)+\ev{\frac{2a}{s} + \vb*{b}\cdot\vb*{\chi}(s,\ell)}.
\end{align}
We can now pose the following SDP problem, 
\begin{align}
\text{Minimize:}\qquad&a,\nonumber\\
    \text{While maintaining:}\qquad&
        \frac{2a-1}{m_3^2} + \vb*{b}\cdot\vb*{\chi}\left(m_3^2,4\right)\geq0\nonumber\\
        &\frac{2a}{m_i^2} + \vb*{b}\cdot\vb*{\chi}\left(m_i^2,\ell\right) \geq0\quad\forall\,\left(i,\ell\right)\neq(3,4),\nonumber\\
        &\frac{2a}{s} + \vb*{b}\cdot\vb*{\chi}(s,\ell) \geq0\quad\forall\,s>M^2,\quad\forall\,\ell\,{\rm even}.\label{eq:sdpbSara}
\end{align}
Fig.~\ref{fig:2state} is obtained in the same way, but  with the ansatz \eqq{eq: RhoDef3}.

In addition to the bound, we can extract the spectrum above $M$ with a method described in Ref.~\cite{Simmons-Duffin:2016wlq}, through the program \texttt{spectrum} from the \texttt{sdpb} repository.
We employ a `threshold' option of $10^{-5}$ to determine whether the polynomials from the primal solution contain a zero or just a minimum at a given value. The couplings below
$M^2$ are extracted directly from the primal solution provided by \texttt{sdpb}  for the $\lambda_{i,\ell}$ in Eq.~\ref{eq:sdpbSara}, up to a normalizing factor.

The parameters used for Figs.~\ref{fig:2state}--\ref{fig:3state_muChange} (and
everywhere else unless specified otherwise) are:
\begin{itemize}
    \item We impose positivity over all even spins $\ell\in[0,500]$,
    as well as additional values of $\ell=500\times10^\kappa$, where $\kappa=3,4,5$, and the  $\ell\to\infty$ limit.
    \item We include every null constraint (Eq.~\eqref{eq:NCS}), with $n\in[3,18]$.
    \item We use a floating-point precision of $2048$ bits in every step of the process, and we increase the maximum allowed number of iterations of the solver from $500$ to $5000$.
\end{itemize}

\subsection{Convergence}\label{app:conv}

\begin{figure}
    \centering
    \includegraphics[width=0.9\textwidth]{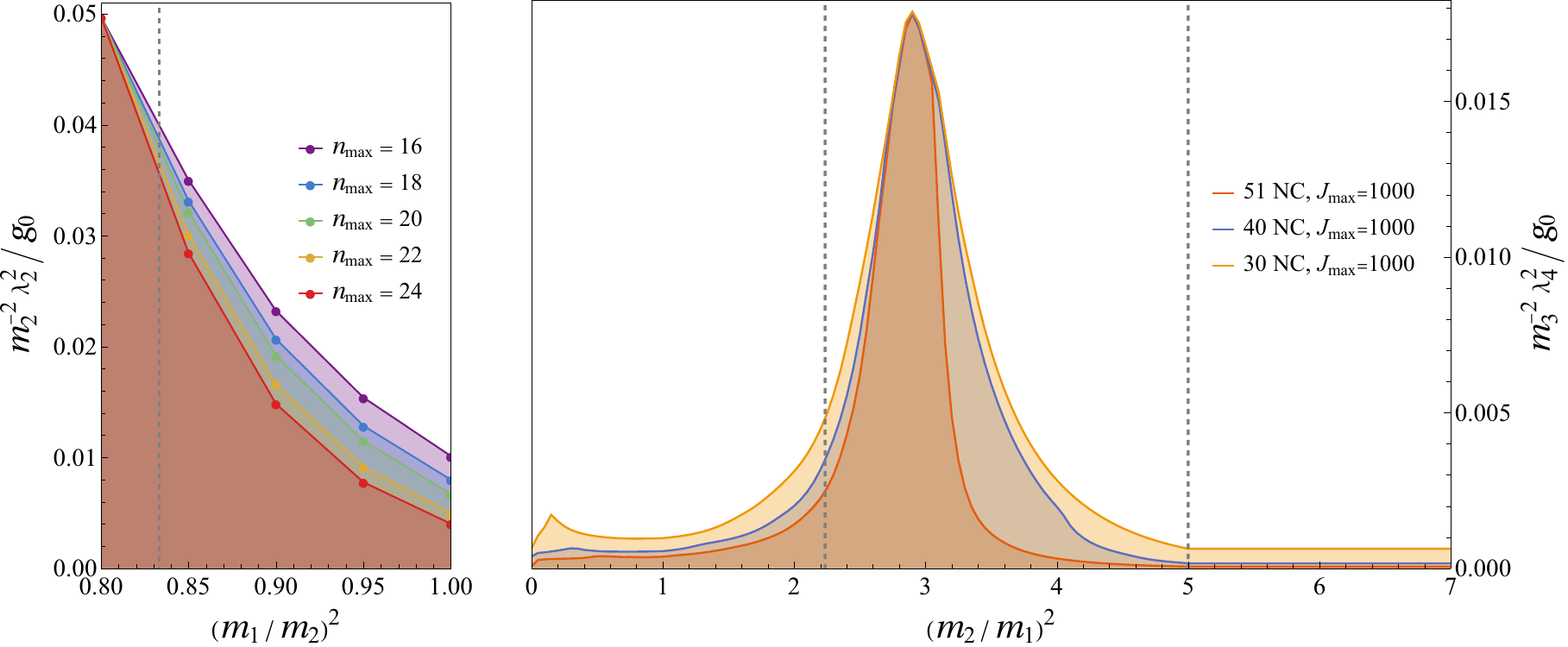}
    \caption{\footnotesize\emph{
    LEFT: Bound on $m_2^{-2}\lambda_{2,2}/g_0$ as a function of $m_1^2$,
    for $M^2=1.2m_2^2$, and for different values of $n_{\rm max}$, the maximum value of $n$ in Eq.~\eqref{eq:NCS}.
    RIGHT: Bound on $m_3^{-2}\lambda_{3,4}/g_0$ as a function of $m_2^2$,
    for $m_3^2=5m_1^2$, $M^2=7m_1^2$, and for different numbers of null constraints.
    These bounds are computed with a different setup than the one used for the left panel: all even spins are included up to $1000$,
    as well as $\ell=2000,3000,5000$ and $\ell\to\infty$. The gray dashed lines indicate the analytic upper and lower bounds~\cite{Berman:2024owc,Berman:2024kdh}.
    }}\label{fig:NCConv}
\end{figure}

In Fig.~\ref{fig:2state} and Fig.~\ref{fig:3state_muChange},
we see tails that extend beyond the analytic constraints from Refs.~\cite{Berman:2024kdh,Berman:2024owc}.
We study this in more detail in Fig.~\ref{fig:NCConv}. The left panel   shows convergence of the tail on the $M^2=1.2m_2^2$ curve from Fig.~\ref{fig:2state},
 by increasing the number of null constraints, for fixed spin truncation. We see that the critical point is stable, while the values on the tail consistently decrease the
more null constraints we add. The right panel  shows the same behavior, but for 3 states below threshold:
tails converge, while the peak remains unaffected by the increase in the number of null constraints.

\subsection{Spectra for the 3-state Problem}\label{app:3st}

\begin{figure}
    \centering
    \includegraphics[width=0.9\textwidth]{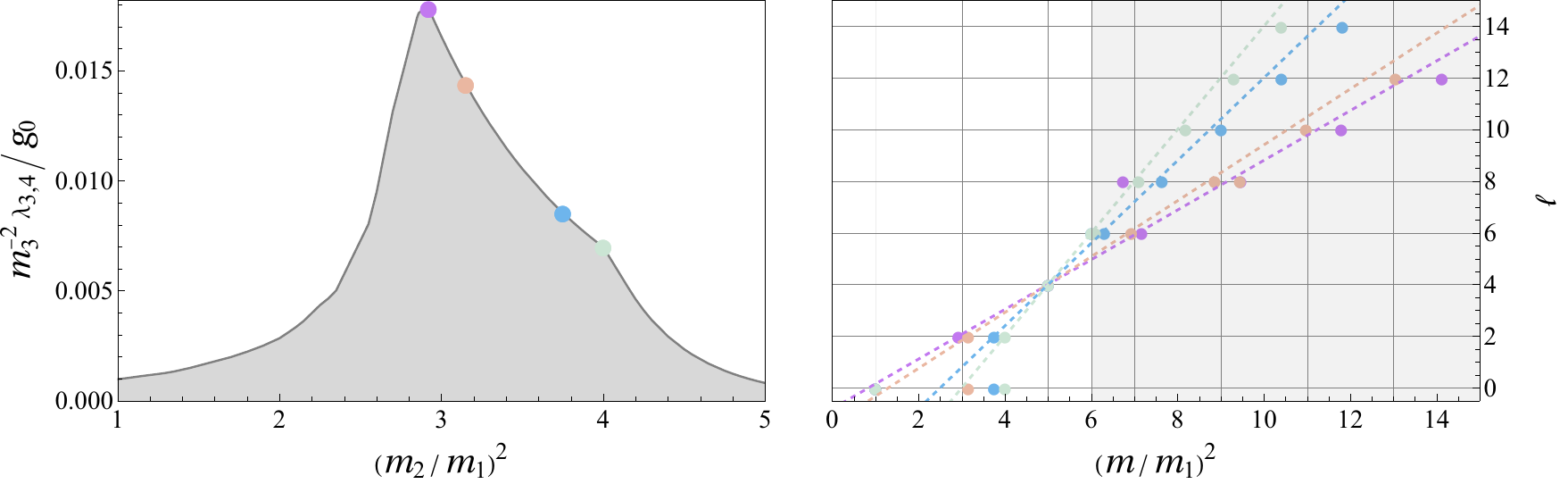}
    \caption{\footnotesize\emph{LEFT: bound from Fig.~\ref{fig:3state_muChange} for $M^2=6m_1^2$ with selected  points highlighted for reference to the right panel. RIGHT:
     spectra corresponding to the highlighted points (only
    $\lambda_{i,\ell}\geq10^{-5}g_0m_1^2$  shown). In shaded gray, the UV region above $M^2$.
    Dashed lines show  slopes predicted by the values of $m_2$ and $m_3$ for each spectrum.}}\label{fig:spectra3States}
\end{figure}

Fig.~\ref{fig:spectra3States} shows the optimal spectra for a selection of points taken on the purple curve in Fig.~\ref{fig:3state_muChange}. For values of $m_2^2$
above the peak, and up to the kink, we  observe that all states---with the exception of $m_1^2$---lie on linear trajectories, which appear in the IR and extend  in the UV beyond $M^2$. 
For these trajectories, the coupling of the spin-0 at $m_2$ (part of a sub-leading trajectory) is not suppressed.

\begin{figure}
    \centering
    \includegraphics[width=0.5\textwidth]{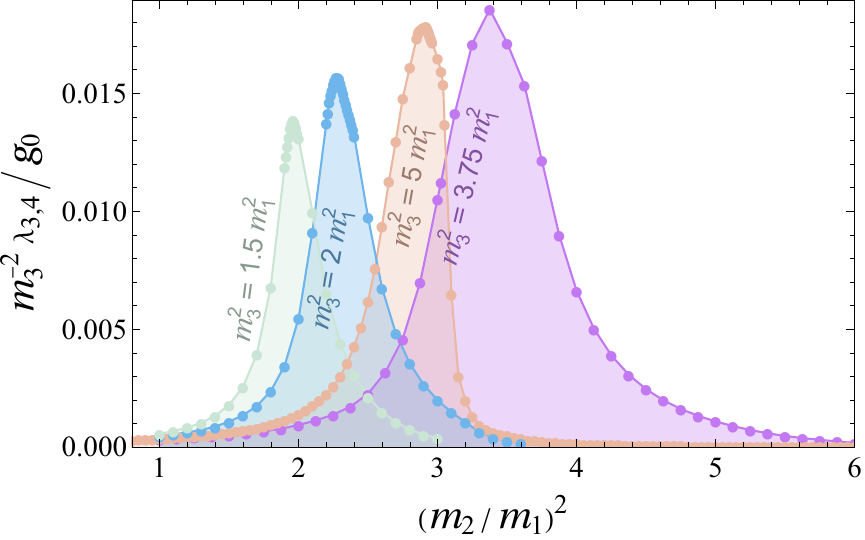}
    \caption{\footnotesize\emph{Bound on $m_3^{-2}\lambda_{3,4}/g_0$ as a function of $m_2^2$
    for different choices of $m_3$, with $M^2$  critical, i.e.\ lying on the linear trajectory defined by $(m_1^2,m_3^2)$. The bound is computed with even spins up to $\ell=1000$, except for $m_3^2=3.75m_1^2$, where $\ell$ goes up to $10000$.}}\label{fig:3state_slope}
\end{figure}

Fig.~\ref{fig:3state_muChange} is computed assuming
the particular relation $m_3^2=5m_1^2$. In Fig.~\ref{fig:3state_slope}, we show the same bound computed
for different slopes. The height of the peak is not significantly affected by this choice. The position of the peak falls very close to $m_2^2=(m_1^2+m_3^2)/2$ for each of the curves, indicating that the results found in Fig.~\ref{fig:3state_muChange} are  generic and that the bounds exhibit no particular preference for the slope, as long as a linear trajectory is formed.

\section{Maximal SUSY Numerical Implementation}\label{app:jedi_appendix}

Gravity, and the associated graviton pole, is unavoidable in Maximal SUSY, and appears in the dispersion relation for $-2$ subtractions,
\begin{eqnarray}
-\frac{8\pi G}{t} = \left\langle (2m^2{+}t)P_J\left(1{+}\frac{2t}{m^2}\right)\right\rangle,
\end{eqnarray}
where the averege is defined in \eqq{eq:fixedt}. 
Because of the pole, we can not take the  forward limit and must use dispersion relations at $t\neq0$~\cite{Caron-Huot:2021rmr}.
We do this as follows.
We use our ansaetze for the spectrum   \eqq{eq: RhoDef2} and \eqq{eq: RhoDef3} to obtain,
\begin{align}
-\frac{8\pi G}{t} &= \sum_{i=1}^{k}\sum_{\text{even}\, J=0}^{2i-2}\lambda_{i,J}\left(2m_i^2{+}t\right)P_J\left(1{+}\frac{2t}{m_i^2}\right){+}\left\langle (2m^2{+}t)P_J\left(1{+}\frac{2t}{m^2}\right)\right\rangle,
\end{align}
where $k=2(3)$ for two (three) states below threshold.

We  then smear  both sides of this relation against a generic functional. 
Following Ref.~\cite{Albert:2024yap} we write,
\begin{eqnarray}
8\pi G \vec{v}_{\text{obj}}{+}\lambda_\phi \vec{v}_{\text{norm}}{+}\sum_{i=1}^{k}\sum_{\text{even}\, J=0}^{2i-2}\lambda_{m_i,J}\vec{v}_{\text{HE}}(m_i^2, J){+}\left\langle \vec{v}_{\text{HE}}(m^2, J)\right\rangle = 0,
\end{eqnarray}
where,
\begin{eqnarray}
\vec{v}_{\text{obj}} &=& (2m_1^{1/2}, \cdots, 0, \cdots)\nonumber\\
\vec{v}_{\text{HE}}(m^2, J) &=& 
\begin{cases}
0 &\text{if } m = m_\phi\, \text{and}\, J = J_\phi \\
(-G_1(m^2, J), \cdots, \mathcal{X}_{1,0}(m^2, J), \cdots) & \text{otherwise}
\end{cases}
\nonumber\\
\vec{v}_{\text{norm}} &=& (-G_1(m_\phi^2, J_\phi), \cdots, \mathcal{X}_{1,0}(m_\phi^2, J_\phi), \cdots)
\end{eqnarray}
with $G_{k}(m^2, J)$ and $\mathcal{X}_{k,q}(m, J)$ defined in Ref.~\cite{Albert:2024yap}, and $\lambda_\phi$  the coupling  that we maximise. For example,  to bound the spin-2 state coupling at mass $m_2^2$, $J_\phi = 2, m_\phi^2 = m_2^2$, while in the three-states system, $J_\phi = 4, m_\phi^2 = m_3^2$. 

We then search for a vector $\bm{\alpha}$  maximizing the objective $\bm{\alpha} \cdot \vec{v}_{\text{obj}}$ subject to the constraints,
\begin{eqnarray}
\bm{\alpha} \cdot \vec{v}_\text{norm} = 1,\quad\bm{\alpha} \cdot \vec{v}_{\text{HE}}(m^2, J)\geq 0,
\end{eqnarray}
for $(m^2, J) \in \mathcal{S} := \{\, (m^2,J) \;\mid\; (m^2 = m_n^2 \wedge J = J_n)\;\lor\;(m^2 > M^2 \wedge J \in 2\mathbb{Z}) \,\}$, where different $m_n, J_n$ correspond to the isolated states  below $M$.

\begin{figure}[H]
\centering
\includegraphics[width = 0.6\linewidth]{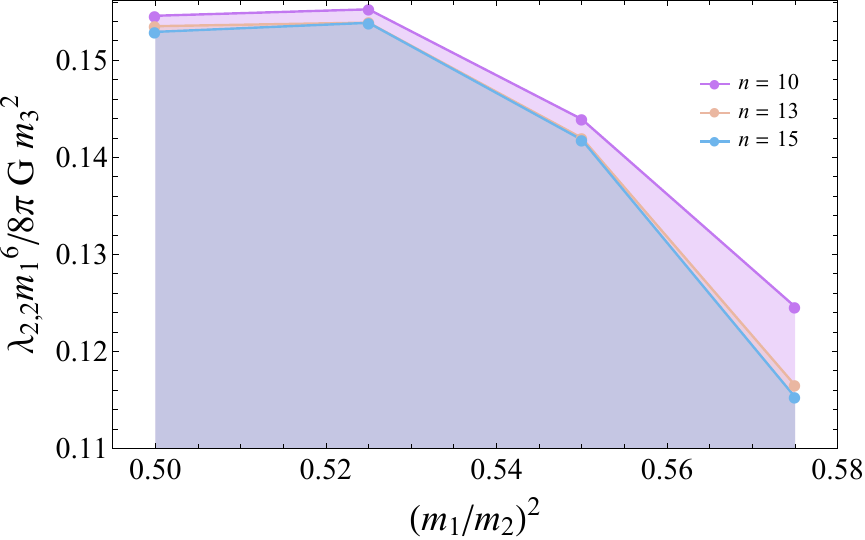}
\caption{\footnotesize\emph{Convergence of the results of Fig.~\ref{fig:2stateTheory}, as a function of the number of null constraints $n$.}
}\label{null_conv_grav}
\end{figure}

We follow a similar implementation as Ref.~\cite{Albert:2024yap}, including small/large/huge values of  $J$, small $m$, and small/large $b$.
\begin{table}[h]
\begin{tabular}{c|c}
params          & value \\ \hline
$J_\text{max}$  & 40    \\
$J_\text{huge}$ & 5000  \\
$m_\text{max}$  & 10    \\
$b_\text{max}$  & 80 \\
$n_\text{max}$ & 13 \\
$k_\text{max}$ & 10 \\
\end{tabular}
\end{table}
We have tested dependence of our results on $J_{\text{huge}}$ up to 10000 and $m_{\text{max}}$ up to 50, and found good convergence. We have also tested convergence in the number of null constraints $n$, which we illustrate in Fig.~\ref{null_conv_grav}, showing a zoomed-in version of Fig.~\ref{fig:2stateTheory}, which exhibits good convergence in proximity to the peak. With more null constraints, however, the computation time for \texttt{sdpb} increases significantly; therefore, many of the plots in this work use $k_\text{max} = 10$.

\pagenumbering{alph}

\twocolumngrid{}
\bibliography{refs}

\end{document}